\documentclass[11pt,a4paper]{article}
\usepackage{jheppub}
\usepackage[T1]{fontenc}
\usepackage{amssymb}
\usepackage{mathtools}
\usepackage{stackengine}
\usepackage{scalerel}

\usepackage{hyperref}
\usepackage{epsfig}
\usepackage{amsmath,latexsym,amssymb}
\usepackage{graphicx}
\usepackage{braket}
\usepackage{pdfsync}

\usepackage{framed}
\usepackage{mathtools}

\usepackage{jheppub}
\usepackage[T1]{fontenc}
\usepackage{amssymb}
\usepackage{mathtools}
\usepackage{amsmath}
\usepackage{bm}
\usepackage{stackengine}

\usepackage{scalerel}
\usepackage{eufrak}
\usepackage{framed}

\usepackage{mathtools}

\usepackage{pstricks}

\usepackage{amsmath}
\usepackage{bbold}
\usepackage{bm}
\usepackage{latexsym}
\usepackage{braket}
\usepackage{slashed}
\usepackage{graphicx,booktabs,multirow}
\usepackage{eufrak}
\numberwithin{equation}{section}

\usepackage{color}
\usepackage{pstricks}
\usepackage{amssymb}

\makeatletter
\newcommand{\doublewidetilde}[1]{{%
  \mathpalette\double@widetilde{#1}%
}}
\newcommand{\double@widetilde}[2]{%
  \sbox\z@{$\m@th#1\widetilde{#2}$}%
  \ht\z@=.9\ht\z@
  \widetilde{\box\z@}%
}
\makeatother

\usepackage{hyperref}
\usepackage{slashed}
\usepackage{epsfig}
\usepackage{amsmath,latexsym,amssymb}
\usepackage{graphicx}
\usepackage[latin1]{inputenc}
\usepackage{braket}
\usepackage{pdfsync}

\usepackage{framed}

\usepackage{mathtools}

\def\be{\begin{equation}}
\def\ee{\end{equation}}
\def\ba{\begin{eqnarray}}
\def\ea{\end{eqnarray}}

\newcommand{\bz}{\bar{z}}

\newcommand{\bh}{\bar{h}}

\newcommand{\bx}{\bar{x}}

\newcommand{\comment}[1]{}

\newcommand{\eea}{\end{eqnarray}}

\author{
Stephan Stieberger${}^{1}$,
Tomasz R.\ Taylor${}^{2}$,\, Bin Zhu${}^2$\\[0.5cm]
$^1${\it Max--Planck--Institut f\"{u}r Physik,	Werner--Heisenberg--Institut, \\80805 M\"unchen, Germany}\\
 $^2${\it Department of Physics \\
  Northeastern University, Boston, MA 02115, USA}\\[0.2cm]
}

\emailAdd{stephan.stieberger@mpp.mpg.de}
\emailAdd{taylor@neu.edu}
\emailAdd{zhu.bi@northeastern.edu}

\title{\boldmath \centerline{Celestial Liouville Theory for Yang-Mills Amplitudes} \unboldmath}
\abstract{We consider Yang-Mills theory with the coupling constant and theta angle determined by the vacuum expectation values of a dynamical (complex) dilaton field. We discuss the tree-level $N$-gluon MHV  scattering amplitudes  in the presence of a nontrivial background dilaton field and construct the corresponding celestial amplitudes by taking Mellin transforms with respect to the lightcone energies. In this way, we obtain two-dimensional CFT correlators of primary fields on the celestial sphere.
We show that the celestial Yang-Mills amplitudes evaluated in the presence of a spherical dilaton shockwave
are given by the correlation functions of primary field operators factorized into the holomorphic current operators times the ``light'' Liouville operators. They are evaluated in the semiclassical limit of Liouville theory (the limit of infinite central charge) and are determined by the classical Liouville field describing metrics on the celestial sphere.\vskip 6cm
\bigskip\bigskip\bigskip}

\notoc
\makeatletter
\gdef\@fpheader{}
\makeatother

\begin{document}
\maketitle
\noindent \section{}\vskip -1cm
Understanding the symmetries is very important for studying physical systems. Typically, the bigger symmetry the better because many problems can be solved by pure symmetry considerations. Celestial holography \cite{Strominger:2017zoo,Raclariu:2021zjz,Pasterski:2021rjz} is a notable exception because  conformal correlators associated with the scattering amplitudes are overconstrained by translational invariance, namely the positions of two-dimensional celestial conformal field theory (CCFT) operators associated with massless particles are constrained by the momentum conservation law. Three-point amplitudes are the extreme case because the operators are driven to the same point on the celestial sphere.\footnote{This problem can be avoided by changing the signature of four-dimensional metrics. This leads, however, outside the scope of a ``garden variety'' Euclidean CCFT in which $SL(2,\mathbb{C})$ conformal symmetry follows directly from Lorentzian $SO(1,3)$.} Similar constraints persist for any number of external particles and complicate the analysis of CCFT correlators \cite{Mizera:2022sln}.

In two recent papers \cite{Fan:2022vbz,Fan:2022kpp}, we analyzed celestial Yang-Mills amplitudes evaluated in the presence of a background dilaton field,  which breaks translational invariance in a controllable way by supplying external momentum to the gluon system. In this theory, the Yang-Mills coupling constant and theta angle are determined by the vacuum expectation values of a dynamical (complex) dilaton field \cite{Dixon:2004za}. A similar approach was pursued in \cite{Costello:2022wso,Casali:2022fro}, in the framework of self-dual gauge theories coupled to axions. In \cite{Fan:2022kpp}, we observed that the four-point correlators factorize into the ``current'' part and the ``scalar'' part. The current factor is given by the group-dependent part of the Wess-Zumino-Witten correlator of four holomorphic currents. The scalar factor can be expressed in terms of  complex integrals of the Dotsenko-Fateev form \cite{Dotsenko:1984ad,Dotsenko:1984nm},
similar to the integrals describing four-point correlators in Coulomb gas models \cite{dots, DiF} and, more generally, in the infinite central charge limit of Liouville theory \cite{Dorn:1994xn,Zamolodchikov:1995aa}.\footnote{Infinite central charge of CCFT was anticipated earlier in \cite{Pasterski:2022lsl}, but on different grounds.} In this Letter, we elaborate on the connection between celestial Yang-Mills amplitudes and Liouville theory.

We begin with a brief summary of Liouville field theory (LFT), along the lines of \cite{Zamolodchikov:1995aa}. The Lagrangian density is given by
\be{\cal L}= \frac{1}{\pi}{\partial \phi\over\partial z}{\partial\phi\over\partial\bz}+\mu e^{2b\phi}\ ,\ee
where $b$ is the dimensionless Liouville coupling constant and $\mu$ is the ``cosmological constant'' scale parameter.
The ``background charge at infinity,''
\be Q=b+\frac{1}{b},\ee
is related to the central charge by
\be c=1+6Q^2.\label{cch}\ee
The primary fields of LFT are the exponential operators
\be V_\alpha (z,\bz)=e^{2\alpha\phi(z,\bz)},\label{lops}\ee
which are scalar fields with chiral dimensions
\be h(\alpha)=\bh(\alpha)=\alpha(Q-\alpha).\ee

The studies of LFT focus on the correlation functions of exponential operators. Their general properties were first analyzed in \cite{Dorn:1994xn,Zamolodchikov:1995aa}, where exact analytic expressions for the three-point functions were also derived. In general, LFT correlation functions are given by complicated expressions. There is, however, one notable exception, for a special configuration of the exponents, when
\be\sum_i\alpha_i=Q-\frac{m}{b}-nb\ .\ee
In this case, the correlation functions become
\begin{align}\nonumber \int &\prod_i V_{\alpha_i}(z_i,\bz_i)\,e^{-\int d^2z\, {\cal L}[\phi]}\, {\cal D}\phi\
\\ &\sim \int\Big\langle\prod_i  V_{\alpha_i}(z_i,\bz_i)
\prod_{k=1}^n  V_{b}(u_k,\bar u_k)\prod_{l=1}^{m}  V_{1/b}(v_l,\bar v_l)
\Big\rangle \,d^2u_1\dots d^2u_n\, d^2v_1\dots d^2v_m\ ,
\end{align}
with the integrands determined by
\begin{align} \Big\langle & \prod_j  V_{\beta_j}(z_j,\bz_j)
\Big\rangle=\prod_{j<j'}|z_j-z_{j'}|^{-4\beta_j\beta_{j'}}\ .
\end{align}
We are interested in the case of single integrals that appear when $n=0,m=1$. With the exponents parametrized as
\be \alpha_i=\sigma_ib\ ,\qquad \sum_i\sigma_i=1\ ,\label{sigmas}\ee
the correlation functions are
\begin{align} \int \prod_i V_{\sigma_ib}(z_i,\bz_i)\,e^{-\int d^2z\, {\cal L}[\phi]}\, {\cal D}\phi\ \sim\, \prod_{i<j}|z_i-z_{j}|^{-4\sigma_i\sigma_jb^2}
\int d^2z \prod_i  |z_i-z|^{-4\sigma_i}\ .\label{intg}
\end{align}
Note that the dimensions of primary fields are
\be d_i=h(\sigma_ib)+\bh(\sigma_ib)=2\sigma_i+2b^2\sigma_i(1-\sigma_i)\ .\label{dims}\ee

We are interested in CCFT primary fields with dimensions
$\Delta_i=n_i+i\lambda_i$, with integer $n_i\leq 1$ and real $\lambda_i$ \cite{Pasterski:2017kqt,Donnay:2020guq}. The integrals encountered in the computations \cite{Fan:2022vbz,Fan:2022kpp} of ``single-valued'' celestial Yang-Mills amplitudes have a form similar to Eq.(\ref{intg}), with $2\sigma_i= m_i+i\lambda_i$ and integer $m_i$. Based on this observation and on Eq.(\ref{dims}), we expect that if the tree-level celestial amplitudes are related to LFT, they are related to the limit $b\to 0$, that is to the $c\to\infty$ infinite central charge (\ref{cch}) limit of LFT. The operators with the exponents scaling as $\sigma_ib$ $(b \to 0)$ are called the ``light'' operators \cite{Zamolodchikov:1995aa}. The limit $b\to 0$ is the classical limit of LFT, in which the correlation functions of light operators are determined by the solutions of the Liouville equation, given by
\be 2b\phi_{0}(z,\bz)=\ln\frac{A}{\pi(1+z\bz)^2}\ \label{classol} \ee
and describing two-dimensional metrics on a sphere with area $A$. The computation of the correlation functions of light operators amounts to
integrating over the $SL(2,\mathbb{C})$ orbits of Eq.(\ref{classol}) in $\prod_i e^{2\sigma_ib\phi_0}$.

In \cite{Fan:2022vbz,Fan:2022kpp}, we computed three- and four-gluon celestial MHV amplitudes in the presence of a specific dilaton background motivated by certain solutions of Banerjee-Ghosh (BG) equations \cite{Banerjee:2020vnt}.
As it is clear, however, from the discussion of Ref.\cite{Hu:2021lrx}, BG equations are satisfied by MHV amplitudes in the presence of an arbitrary dilaton source because the net effect of the source is to replace the momentum-conserving delta functions  by a (Lorentz-invariant) function of the total momentum of the gluon system. The total momentum is invariant under BCFW shifts, therefore BG equations remain satisfied. As mentioned before, the celestial amplitudes obtained in \cite{Fan:2022vbz,Fan:2022kpp} can be expressed in terms of integrals similar to Eq.(\ref{intg}). In this work, we will identify the background dilaton field that yields celestial amplitudes in {\it exactly\/} the same form as the Liouville integrals
(\ref{intg}). In this way, we will connect the operators representing gluons with the Liouville operators (\ref{lops}) and relate CCFT to LFT.

Celestial amplitudes are obtained from the momentum space amplitudes by performing Mellin transforms with respect to the energies of external particles \cite{Pasterski:2017ylz}; therefore our first goal is to rewrite the Liouville integrals (\ref{intg}) as Mellin transforms.
Since we are interested in ``mostly plus'' MHV amplitudes in the helicity configurations $(--++\dots)$, it is natural to fulfill the condition (\ref{sigmas}) by
\be \sigma_1=\frac{1+i\lambda_1}{2},\ \sigma_2=\frac{1+i\lambda_2}{2}
,\ ~\sigma_k=\frac{i\lambda_k}{2} ~~(k\geq 3),\ee
with \be\sum_i^N\lambda_i=0\ .\ee
Hence, we focus on the integrals
\begin{align} I_N(z_1,\bz_1,&\dots ,z_N,\bz_N)
=\int d^2z \, |z_1-z|^{-2(1+i\lambda_1)}   |z_2-z|^{-2(1+i\lambda_2)}    |z_3-z|^{-2i\lambda_3}     \dots  \ , \label{intg1}\end{align}
which describe conformal correlators of  scalar primary fields with dimensions $ d_i=2\sigma_i$.
They can be evaluated in many ways \cite{dots}, but for our purposes it is most convenient to follow the formalism described in \cite{Simmons-Duffin:2012juh} and express them in terms of the integrals on a ``Poincar\'e section'' of a four-dimensional embedding space $\mathbb{R}^{3,1}$.

In the embedding  space, the coordinates are parametrized as
$X=(X^+,X^-,X^1,X^2)$,
and the Lorentzian inner product is
\be X\cdot Y=\frac{1}{2}(X^+Y^-+X^-Y^+)-X^1Y^1-X^2Y^2\ .\ee
On the null-cone $X\cdot X=0$,
\be\label{xxs} X=\Big(X^+,\frac{|z|^2}{X^+},\frac{z+\bz}{2},\frac{z-\bz}{2i}\Big)\ ,\quad z\in\mathbb{C}\ .\ee
The Poincar\'e section is constructed by quotienting the null-cone by the rescaling $X\sim \rho X,\ \rho\in\mathbb{R}$. The projective null-cone can be identified with $\mathbb{C}$ by gauge fixing this rescaling, for example by imposing the condition $X^+=1$. The action of $SO(1,3)$ on $\mathbb{R}^{3,1}$ is inherited as $SL(2,\mathbb{C})$ transformations of the complex coordinates $z$. Note that on the Poincar\'e section,
\be |z_1-z_2|^2=2X_1\cdot X_2\ .\ee

By using the embedding space formalism \cite{Simmons-Duffin:2012juh}, the integrals (\ref{intg1}) can be rewritten as
\begin{align}
I_{N} =&\, \frac{1}{2\,\text{Vol GL}(1,\mathbb{R})^+}\int_{\scriptscriptstyle X^++X^->0} d^4X\delta(X\cdot X)\frac{1}{( X_1\cdot X)^{1+i\lambda_1}( X_2\cdot X)^{1+i\lambda_2}\prod_{k=3}^N(X_k\cdot X)^{i\lambda_k}} \nonumber\\[1mm] \label{eq:I4Duffin}
= &\,\frac{{ C}_N}{\text{Vol GL}(1,\mathbb{R})^+}\int_{\scriptscriptstyle X^++X^->0}d^4X\delta(X\cdot X)\\ &\times\int_{\omega_i\geq 0}  e^{iX \cdot(\sum_{i=1}^N\omega_iX_i ) }\, d\omega_1\,  \omega_1^{i\lambda_1}d\omega_2 \,\omega_2^{i\lambda_2}\prod_{k\geq 3}^Nd\omega_k\,\omega_k^{i\lambda_k-1} \ ,\nonumber
\end{align}
with the normalization constant
\be { C}_N
= -\frac{1}{2\,\Gamma(1+i\lambda_1)\Gamma(1+i\lambda_2)\prod_{k\geq 3}^N\Gamma(i\lambda_k)} \, . \label{eq:factorN3}
\ee
The volume factor $[\text{Vol GL}(1,\mathbb{R})^+]^{-1}$ cancels the divergence due to the integration over the
$X\to\rho X$ rescaling ``gauge'' mode. The gauge can be fixed to $X^+=1$, with the associated Fadeev-Popov determinant equal to 1. Then
\begin{align}
I_{N}
= &\, C_N\int d^4X\delta(X\cdot X)\,\delta(X^+-1)\,\theta(X^++X^-)\label{intg2}\\ &~~~~~~~~~~\times\int_{\omega_i\geq 0}  e^{iX \cdot(\sum_{i=1}^N\omega_iX_i ) }\, d\omega_1\,  \omega_1^{i\lambda_1}d\omega_2 \,\omega_2^{i\lambda_2}\prod_{k\geq 3}^Nd\omega_k\,\omega_k^{i\lambda_k-1} \ .\nonumber
\end{align}
The above integral has a form of a multiple Mellin transform, with implicit regulators \cite{Pasterski:2017kqt} omitted here for simplicity. Let us compare it with the Mellin transforms encountered in celestial amplitudes.

The starting point for celestial amplitudes are the momentum space amplitudes with the momenta of massless particles parametrized as
\be P=(P^+,P^-,P^1,P^2)=\omega \Big(1,|z|^2,\frac{z+\bz}{2},\frac{z-\bz}{2i}\Big).\ee
Note that $\omega$ is the energy  in the light-cone frame. Under $SL(2,\mathbb{C})$ Lorentz transformations,
\be z\to \frac{az+b}{cz+d},\qquad\omega\to{\omega}{|cz+d|^2}.\label{sltr}\ee
We are interested in $N$-gluon MHV amplitudes in the mostly plus helicity configuration $(--++\dots)$, evaluated in the presence of a dilaton source $J_\Phi(X)$
\cite{Fan:2022vbz,Fan:2022kpp}, or equivalently, in the presence of a background dilaton field
$\Phi_0(X)=\Box^{-1}J_{\Phi}$.\footnote{We are considering a ``weakly coupled'' source, with one insertion in the amplitude.}  These amplitudes are converted into two-dimensional correlators of primary fields with dimensions $\Delta_i$ by performing Mellin transforms with respect to the light-cone energies. In this way, we obtain \cite{Fan:2022vbz,Fan:2022kpp}:
\be {\cal M}_N(z_1,\bz_1,\dots,z_N,\bz_N|\Delta_1,\dots,\Delta_N) ={\cal J}_N(z_i)\,{\cal S}_N(z_i,\bz_i)\label{mhv1}\ee
where
\be {\cal J}_N(z_i)=\sum_{\pi\in S_{N-2}}f^{a_1a_{\pi(2)}x_1}f^{x_1a_{\pi(3)}x_2}\cdots f^{x_{N{-}3}a_{\pi(N{-}1)}a_{N}}\frac{z_{12}^4}{z_{1\pi(2)}z_{\pi(2)\pi(3)}\cdots z_{N1}},\label{cfactor} \ee
is the holomorphic ``soft'' factor, with $(a_1,a_2,\dots,a_N)$ labeling the gluon group indices.
The Mellin transforms are contained in the ``scalar'' part
\be {\cal S}_N(z_i,\bz_i)= \label{spart}
\int {d^4X}\int_{\omega_i\geq 0}J_{\Phi}(X)\frac{e^{iX\cdot Q}}{Q^2}   d\omega_1\,  \omega_1^{\Delta_1}d\omega_2 \,\omega_2^{\Delta_2}\prod_{k\geq 3}^Nd\omega_k\,\omega_k^{\Delta_k-2},\ee
where $Q$ is the total momentum of the gluon system,
\be Q=\sum_{i=1}^NP_i\ee
and we assumed that all gluons are outgoing, {\em i.e}., they are created by the background dilaton field. We see that, up to the normalization constant, the scalar part (\ref{spart}) matches the Liouville integral (\ref{intg2}) with the dimensions
\be \Delta_1=i\lambda_1=d_1-1 ,\ \Delta_2=i\lambda_2=d_2-1,\  ~\Delta_k=1+i\lambda_k=d_k+1~~(k\geq 3)\ ,\label{dimens}\ee
where $d_i=2\sigma_i$ are the dimensions of Liouville primary fields, provided that we identify the dilaton source as
\be J_{\Phi}(X)=-\square\,[\delta(X\cdot X)\,\delta(X^+-1)\,\theta(X^++X^-)]\ee
and choose a Poincar\'e section, see Eq.(\ref{xxs}), with all $X_i^+=1$, so that
\be P_i=\omega_iX_i\ ,\quad Q=\sum_{i=1}^N\omega_iX_i\ .\ee
The corresponding background field is
\be\Phi_0(X)=\square^{-1}J_{\Phi}(X)=-\delta(X\cdot X)\,\delta(X^+-1)\,\theta(X^++X^-)\ .\label{phisol}\ee
It represents a retarded (outgoing)  spherical  dilaton shockwave. It seems that the wave is frozen at $X^+=1$, which would violate Lorentz invariance. Nevertheless, as it will be clarified below, it gives rise to a covariant conformal correlator. The wave propagation is ``lost'' by projectivizing on the null-cone.

To further analyze Liouville integrals, we rewrite Eq.(\ref{intg2}) as
\begin{align}
I_{N}
= &\, \frac{C_N}{2}\int dX^+dX^-d^2\vec X\,\delta(X^+X^--\vec{X}^2)\,\delta(X^+-1)\,\theta(X^++X^-)\nonumber \\ &~~~~~~~~~~\times\int_{\omega_i\geq 0}  e^{iX \cdot Q}\, d\omega_1\,  \omega_1^{i\lambda_1}d\omega_2 \,\omega_2^{i\lambda_2}\prod_{k\geq 3}^Nd\omega_k\,\omega_k^{i\lambda_k-1} \label{intg3}\\
= &\, \frac{C_N}{2}\int d^2\vec X\, \int_{\omega_i\geq 0}  \exp\Big[\frac{i}{2}(Q^+\vec X^2-2\vec Q\cdot\vec X+Q^- )\Big]  \, d\omega_1\,  \omega_1^{i\lambda_1}d\omega_2 \,\omega_2^{i\lambda_2}\prod_{k\geq 3}^Nd\omega_k\,\omega_k^{i\lambda_k-1}\ .\nonumber
\end{align}
As a result of Gaussian integration, we obtain
\begin{align}
I_{N}
= &\, {C_N} \int_{\omega_i\geq 0} \frac{i\pi}{Q^+} \exp\Big(\frac{iQ^2}{2Q^+}\Big)  \, d\omega_1\,  \omega_1^{i\lambda_1}d\omega_2 \,\omega_2^{i\lambda_2}\prod_{k\geq 3}^Nd\omega_k\,\omega_k^{i\lambda_k-1}\label{intg9}\\ =&\,
{C_N} \int_{\omega_i\geq 0} \frac{i\pi}{\sum_{i=1}^N\omega_i} \exp\Big[\frac{i\sum_{i<j}^N\omega_i\omega_jz_{ij}\bz_{ij}}{2\sum_{i=1}^N\omega_i}\Big]  \, d\omega_1\,  \omega_1^{i\lambda_1}d\omega_2 \,\omega_2^{i\lambda_2}\prod_{k\geq 3}^Nd\omega_k\,\omega_k^{i\lambda_k-1}.\nonumber
\end{align}
It is convenient to change the integration variables to
\be\omega=\sum_{i=1}^N\omega_i,\qquad y_k=\frac{\omega_k}{\sum_{i=1}^N\omega_i}\quad (k=1,\dots, N-1)\ . \ee Then
\begin{align}
I_{N}
 =&\, i\pi
{C_N} \Big(\prod_{k=1}^{N-1}\int_0^1dy_k\Big)\,
y_1^{i\lambda_1}y_2^{i\lambda_2}\prod_{j\geq 3}^{N-1}y_j^{i\lambda_j-1}(1-{\textstyle \sum_{l=1}^{N-1}y_l})^{i\lambda_N-1}
\\ &\times \int_{0}^\infty d\omega \exp\Big[\frac{i\omega}{2}\big(\sum_{i<j}^{N{-}1}y_iy_jz_{ij}\bz_{ij}
+(1-{\textstyle \sum_{l=1}^{N-1}y_l})\sum_{i=1}^{N}y_iz_{iN}\bz_{iN}\big)\Big]  \nonumber\\[1mm]
=&\, -2\pi
{C_N} \Big(\prod_{k=1}^{N-1}\int_0^1dy_k\Big)\,
y_1^{i\lambda_1}y_2^{i\lambda_2}\prod_{j\geq 3}^{N-1}y_j^{i\lambda_j-1}(1-{\textstyle \sum_{l=1}^{N-1}y_l})^{i\lambda_N-1}\nonumber\\ &~~~~~\times
\Big[\sum_{i<j}^{N{-}1}y_iy_jz_{ij}\bz_{ij}
+(1-{\textstyle \sum_{l=1}^{N-1}y_l})\sum_{i=1}^{N}y_iz_{iN}\bz_{iN}\Big]^{-1}\ .\label{intg5}
\end{align}
Recall that the above integral (\ref{intg5}) is evaluated on the support of $\sum_i\lambda_i=0$. In order to make Lorentz covariance explicit, we multiply it by
\be
2\pi\delta(\sum_i\lambda_i)=\int_0^\infty \omega^{(i\sum_{i=1}^N\lambda_i)-1}d\omega\ ,\ee
and return to the original integration variables,
\be \omega_k=y_k\omega\quad (k=1,\dots, N-1)\ ,~~\omega_N=(1-{ \sum_{k=1}^{N-1}y_k})\omega\ .\ee
In this way, we obtain
\be-2\pi  C_N^{-1}\,\delta(\sum_i\lambda_i)\,I_{N}
= \int_{\omega_i\geq 0}\Big(\frac{2\pi}{Q^2}\Big)\,  d\omega_1\,  \omega_1^{i\lambda_1}d\omega_2 \,\omega_2^{i\lambda_2}\prod_{k\geq 3}^Nd\omega_k\,\omega_k^{i\lambda_k-1}\ , \label{intg6}\ee
where
\be
Q^2=\sum_{i<j}^N\omega_i\omega_j z_{ij}\bz_{ij}\ . \ee
The final expression (\ref{intg6}) is manifestly Lorentz covariant under $SL(2,\mathbb{C})$ transformations (\ref{sltr}). It matches the scalar part (\ref{spart})  of celestial amplitudes with the primary field dimensions written in Eq.(\ref{dimens}) and with the scalar current
\be {\cal J}_\Phi (x)=2\pi\delta^{(4)}(X)\ .\label{jfi}\ee
The solution
of the  equation $\Box\Phi_0(X)=2\pi\delta^{(4)}(X)$,
matching the shockwave (\ref{phisol})  on the Poincar\'e section, is  the background field
\be\Phi_0(X)=\Phi_0(r,t)=-\frac{1}{2r}\delta(r-t)\,\theta(t)\ .\label{backg}\ee

As a consistency check, we evaluate Liouville integrals (\ref{intg1}) for $N=3$ and $N=4$ by using our method and compare with the results reported previously in the literature. For $N=3$, we start from Eq.(\ref{intg6}):
\begin{align}\nonumber\delta({\sum_i\lambda_{i}})& I_{3}
= -C_3\int d\omega_1d\omega_2d\omega_3 \, \omega_1^{i\lambda_1} \,\omega_2^{i\lambda_2}\omega_3^{i\lambda_3-1}(\,\omega_1\omega_2 z_{12}\bar{z}_{12}+\omega_2\omega_3 z_{23}\bar{z}_{23}+\omega_1\omega_3 z_{13}\bar{z}_{13})^{-1}\\
=&\,-C_3\int d\omega_1 d\omega_2 \, \omega_1^{i\lambda_1}\omega_2^{i\lambda_2}B(i\lambda_3,1-i\lambda_3)(\omega_1 z_{13}\bar{z}_{13}+\omega_2 z_{23}\bar{z}_{23})^{-i\lambda_3}(\omega_1\omega_2 z_{12}\bar{z}_{12})^{i\lambda_3-1} \nonumber\\
=&\, -C_3 \int d\omega_1 \omega_1^{i\lambda_1+i\lambda_2+i\lambda_3-1}B(i\lambda_3,1-i\lambda_3)B(-i\lambda_2,i\lambda_2+
i\lambda_3)\nonumber \\ &~~~~~~~~~~~~~~~~~\times (z_{12}\bar{z}_{12})^{i\lambda_3-1}(z_{13}\bar{z}_{13})^{i\lambda_2}(z_{23}
\bar{z}_{23})^{-i\lambda_2-i\lambda_3} \nonumber\\[1mm]
=&\, \pi \delta( \sum_i\lambda_i)\,(z_{12}\bar{z}_{12})^{i\lambda_3-1} (z_{23}\bar{z}_{23})^{i\lambda_1} (z_{13}\bar{z}_{13})^{i\lambda_2} \frac{\Gamma(-i\lambda_1)\Gamma(-i\lambda_2)\Gamma(1-i\lambda_3)}{\Gamma(1+i\lambda_1)
\Gamma(1+i\lambda_2)\Gamma(i\lambda_3)}\ .\label{i3fin}
\end{align}
In this way, we obtain
\be I_3= \pi \frac{\Gamma(-i\lambda_1)\Gamma(-i\lambda_2)\Gamma(1-i\lambda_3)}{\Gamma(1+i\lambda_1)
\Gamma(1+i\lambda_2)\Gamma(i\lambda_3)}\,
(z_{12}\bar{z}_{12})^{i\lambda_3-1} (z_{23}\bar{z}_{23})^{i\lambda_1} (z_{13}\bar{z}_{13})^{i\lambda_2} \ ,\label{i3fin}\ee
in agreement with \cite{Zamolodchikov:1995aa,Dolan:2011dv}. For $N=4$, a similar computation yields
\begin{align}
I_4 \label{i4fin}
&=(z_{12}\bar{z}_{12})^{i\lambda_3+i\lambda_4-1} (z_{13}\bar{z}_{13})^{-i\lambda_3} (z_{14}\bar{z}_{14})^{i\lambda_2+i\lambda_3} (z_{24}\bar{z}_{24})^{i\lambda_1} \\[1mm]
&\quad\times \Big[ C_{\Delta} K_{34}^{21}[\textstyle \frac{i\lambda_3}{2}+\frac{i\lambda_4}{2},
\frac{i\lambda_3}{2}+\frac{i\lambda_4}{2}](x,\bar{x}) +C_{\widetilde{\Delta}} K^{21}_{34}[\textstyle 1-\frac{i\lambda_3}{2}-\frac{i\lambda_4}{2},1-\frac{i\lambda_3}{2}
-\frac{i\lambda_4}{2}](x,\bar{x}) \Big] \, , \nonumber\end{align}
where $x$ is the conformal invariant cross ratio,
\be
x=\frac{z_{12}z_{34}}{z_{13}z_{24}} \, ,
\ee
and $K^{21}_{34}[h,\bh](x,\bx)$ are the two-dimensional $s$-channel global conformal blocks  \cite{Dolan:2000ut,Dolan:2003hv,Zamolodchikov:1984eqp,Fan:2021isc,Fan:2021pbp} associated with primary fields with dimensions $\Delta=h+\bh$:
\begin{align}
K^{21}_{34}[h,\bar{h}]&(x,\bar{x}) \\
=&~x^{h-h_3-h_4}\bar{x}^{\bar{h}-\bar{h}_3-\bar{h}_4} \,_2F_1\bigg({h-h_{12},h+h_{34}\atop 2h};x\bigg) \,_2F_1\bigg({\bar{h}-\bar{h}_{12},\bar{h}+\bar{h}_{34}\atop 2\bar{h}};\bar{x}\bigg) \, , \nonumber
\end{align}
where $h_{ij}=h_i-h_j, ~\bh_{ij}=\bh_i-\bh_j$. In the case under consideration, $h_i=\bh_i$ and $h=\bh$.
We find one block with dimension $\Delta=i\lambda_3+i\lambda_4$ and its shadow block with dimension $\widetilde\Delta=2-\Delta$, with the coefficients
\begin{align}
C_{\Delta} &= \pi\frac{\Gamma(-i\lambda_1)
\Gamma(-i\lambda_2)\Gamma(1+i\lambda_1+i\lambda_2)}{\Gamma(1+i\lambda_1)
\Gamma(1+i\lambda_2)\Gamma(-i\lambda_1-i\lambda_2)} \, ,\\
C_{\widetilde{\Delta}} &=\pi \frac{\Gamma(1-i\lambda_3)\Gamma(1-i\lambda_4)
\Gamma(i\lambda_3+i\lambda_4-1)}{\Gamma(i\lambda_3)
\Gamma(i\lambda_4)\Gamma(2-i\lambda_3-i\lambda_4)} \, .
\end{align}
Eq.(\ref{i4fin}) agrees with the expression written in Ref.\cite{Simmons-Duffin:2012juh}. Furthermore, we checked that this four-point correlator satisfies the crossing symmetry constraints. For instance, in the $u$-channel, in which $x\to 1-x$, one also finds a single block with dimension
$\Delta=1+i\lambda_2+i\lambda_3$ and a shadow block.

Now we turn to the  ``soft'' part (\ref{cfactor}) of celestial MHV amplitudes. Here, the negative helicity $-1$ gluons can be associated with the holomorphic operators $\widehat J^a(z)$ in the adjoint representation of the gauge group, with chiral weights $(h=-1,\bh=0)$, {\em i.e}., dimensions $-1$, while the positive helicity $+1$ gluons to dimension $+1$ holomorphic $(h=1,\bh=0)$ Wess-Zumino-Witten (WZW) currents $J^a(z)$.\footnote{Since the dimension of $\widehat J^a(z)$ is negative, it is not clear how to incorporate this operator into the framework of unitary WZW models.} The correlators involving  WZW currents are completely determined by Ward identities, which, in the context of celestial holography, follow from the leading soft gluon theorem
\cite{Strominger:2017zoo,Raclariu:2021zjz,Pasterski:2021rjz}. Assuming
\be \langle \widehat J^{a_1}(z_1) \widehat J^{a_2}(z_2)\rangle=\delta^{a_1a_2}z_{12}^2\ ,\ee
one obtains \cite{Costello:2022wso}:
\be \langle \widehat J^{a_1}(z_1)  \widehat J^{a_2}(z_2)
J^{a_3}(z_3) \cdots J^{a_N}(z_N)\rangle={\cal J}_N(z_i)\ .\ee

At this point, we are in a position to connect the tree-level celestial MHV amplitudes to the classical limit of  Liouville theory. We introduce the following operators:
\begin{align} {\cal O}^{-a}_{\lambda}(z,\bz)& ~=~\Gamma(1+i\lambda)\, \widehat J^a(z)\,e^{(1+i\lambda ) b\phi(z,\bz)}\\[2mm]
 {\cal O}^{+a}_{\lambda}(z,\bz)& ~=~\Gamma(i\lambda)\, J^a(z)\,e^{i\lambda b\phi(z,\bz)}
\end{align}
and consider the limit of $b\to 0$. In this limit, the dimension of ${\cal O}^{-a}_{\lambda}$ becomes $\Delta_-=i\lambda$ while the dimension of ${\cal O}^{+a}_{\lambda}$ is $\Delta_+=1+i\lambda$.
We associate ${\cal O}^{\pm a}_{\lambda}$ to gluons with helicities $\pm 1$, respectively, and group indices $a$. From our discussion, {\em c.f}.\ Eqs.(\ref{intg6}) and (\ref{eq:factorN3}), it follows that
\begin{align}
\label{fin1}
4\pi\delta\big(\sum_{i=1}^{N}\lambda_i\big)\,&\Big\langle{\cal O}^{-a_1}_{\lambda_1}(z_1,\bz_1)\,
{\cal O}^{-a_2}_{\lambda_2}(z_2,\bz_2)\,
{\cal O}^{+a_3}_{\lambda_3}(z_3,\bz_3)\cdots
{\cal O}^{+a_N}_{\lambda_N}(z_N,\bz_N)\Big\rangle\\ \nonumber &~~~~~=\,{\cal M}_N
(z_1,\bz_1,\dots,z_N,\bz_N|i\lambda_1,i\lambda_2,1+i\lambda_3,\dots,1+i\lambda_N)  ,
\end{align}
where ${\cal M}_N$ is the celestial MHV amplitude, {\em c.f}.\ Eqs.(\ref{mhv1}-\ref{spart}),
evaluated in the shockwave background of Eqs.(\ref{jfi},\ref{backg}). Note that the dimensions of negative helicity gluons are shifted by $-1$ as compared with the conventional ``principal series'' celestial amplitudes \cite{Pasterski:2017ylz}. By using Eqs.(\ref{i3fin}) and (\ref{i4fin}), we checked explicitly that the correlators (\ref{fin1}) satisfy BG equations for $N=3$ and $N=4$. We also checked that the operator product expansions extracted from the collinear limits of four-gluon amplitudes agree with Refs.\cite{Fan:2019emx,Pate:2019lpp}

The fact that a two-dimensional spherical shell of the dilaton shockwave propagating in four-dimensional spacetime corresponds to the classical Liouville field describing metrics on the celestial sphere is perhaps the most natural, not to say expected, outcome of our analysis.
Liouville theory has been linked before to Yang-Mills through the AGT correspondence
\cite{Alday:2009aq}, which raises the question whether our ``dilaton deformation'' is related in some way to Nekrasov's deformation \cite{Nekrasov:2002qd,Nekrasov:2003rj}.\footnote{Although AGT correspondence applies to superconformal Yang-Mills theory, there is no difference between supersymmetric and non-supersymmetric gluon amplitudes at the tree level \cite{Parke:1985pn,Taylor:2017sph}.}
It would also be very interesting to connect Liouville shockwaves to the shockwaves recently studied in \cite{deGioia:2022nkq,Gonzo:2022tjm}.
The most interesting and possibly related problem, however, is to understand how the quantum corrections to celestial Yang-Mills amplitudes are related to quantum Liouville theory with finite central charge. Answering this question would create a novel path towards Yang-Mills dynamics.\\[2ex]\noindent {\bf Acknowledgments}\\[2mm]
We thank Wei Fan and Angelos Fotopoulos for fruitul collaborations leading to the present work.
We are grateful to Daniel Kapec, Sabrina Pasterski, Andy Strominger and Herman Verlinde  for helpful discussions.
This material is based in part upon work supported by the National Science Foundation
under Grants Number PHY--1913328 and PHY--2209903.
Any opinions, findings, and conclusions or recommendations
expressed in this material are those of the authors and do not necessarily
reflect the views of the National Science Foundation.

\notoc


\begin{thebibliography}{99}
\bibitem{Strominger:2017zoo}
  A.~Strominger,
  {\it Lectures on the Infrared Structure of Gravity and Gauge Theory},
Princeton University Press (2018)
  [arXiv:1703.05448 [hep-th]].
\bibitem{Raclariu:2021zjz}
A.~M.~Raclariu,
``Lectures on Celestial Holography,''
[arXiv:2107.02075 [hep-th]].
\bibitem{Pasterski:2021rjz}
S.~Pasterski,
``Lectures on Celestial Amplitudes,''
Eur. Phys. J. C \textbf{81} (2021) no.12, 1062
[arXiv:2108.04801 [hep-th]].
\bibitem{Mizera:2022sln}
S.~Mizera and S.~Pasterski,
``Celestial Geometry,''
[arXiv:2204.02505 [hep-th]].
\bibitem{Fan:2022vbz}
W.~Fan, A.~Fotopoulos, S.~Stieberger, T.~R.~Taylor and B.~Zhu,
``Elements of celestial conformal field theory,''
JHEP \textbf{08}, 213 (2022)
doi:10.1007/JHEP08(2022)213
[arXiv:2202.08288 [hep-th]].
\bibitem{Fan:2022kpp}
W.~Fan, A.~Fotopoulos, S.~Stieberger, T.~R.~Taylor and B.~Zhu,
``Celestial Yang-Mills Amplitudes and D=4 Conformal Blocks,''
[arXiv:2206.08979 [hep-th]].
\bibitem{Dixon:2004za}
L.~J.~Dixon, E.~W.~N.~Glover and V.~V.~Khoze,
``MHV rules for Higgs plus multi-gluon amplitudes,''
JHEP \textbf{12} (2004), 015
doi:10.1088/1126-6708/2004/12/015
[arXiv:hep-th/0411092 [hep-th]].
\bibitem{Costello:2022wso}
K.~Costello and N.~M.~Paquette,
``Celestial holography meets twisted holography: 4d amplitudes from chiral correlators,''
[arXiv:2201.02595 [hep-th]].
\bibitem{Casali:2022fro}
E.~Casali, W.~Melton and A.~Strominger,
``Celestial Amplitudes as AdS-Witten Diagrams,''
[arXiv:2204.10249 [hep-th]].
\bibitem{Dotsenko:1984ad}
V.~S.~Dotsenko and V.~A.~Fateev,
``Four Point Correlation Functions and the Operator Algebra in the Two-Dimensional Conformal Invariant Theories with the Central Charge c \ensuremath{<} 1,''
Nucl. Phys. B \textbf{251}, 691-734 (1985)
doi:10.1016/S0550-3213(85)80004-3
\bibitem{Dotsenko:1984nm}
V.~S.~Dotsenko and V.~A.~Fateev,
``Conformal Algebra and Multipoint Correlation Functions in Two-Dimensional Statistical Models,''
Nucl. Phys. B \textbf{240}, 312 (1984)
doi:10.1016/0550-3213(84)90269-4
\bibitem{dots} V. Dotsenko, ``S\'erie de Cours sur la Th\'eorie Conforme''. DEA. 2006. <cel-00092929>. https://cel.archives-ouvertes.fr/cel-00092929

\bibitem{DiF} P. Di Francesco, P. Mathieu, D. S\'en\'echal, ``Conformal Field Theory,'' Springer (1997).
\bibitem{Dorn:1994xn}
H.~Dorn and H.~J.~Otto,
``Two and three point functions in Liouville theory,''
Nucl. Phys. B \textbf{429}, 375-388 (1994)
doi:10.1016/0550-3213(94)00352-1
[arXiv:hep-th/9403141 [hep-th]].
\bibitem{Zamolodchikov:1995aa}
A.~B.~Zamolodchikov and A.~B.~Zamolodchikov,
``Conformal bootstrap in Liouville field theory,''
Nucl. Phys. B \textbf{477}, 577-605 (1996)
doi:10.1016/0550-3213(96)00351-3
[arXiv:hep-th/9506136 [hep-th]].
\bibitem{Pasterski:2022lsl}
S.~Pasterski and H.~Verlinde,
``Chaos in Celestial CFT,''
[arXiv:2201.01630 [hep-th]].
\bibitem{Pasterski:2017kqt}
S.~Pasterski and S.~H.~Shao,
``Conformal basis for flat space amplitudes,''
Phys. Rev. D \textbf{96}, no.6, 065022 (2017)
doi:10.1103/PhysRevD.96.065022
[arXiv:1705.01027 [hep-th]].
\bibitem{Donnay:2020guq}
L.~Donnay, S.~Pasterski and A.~Puhm,
``Asymptotic Symmetries and Celestial CFT,''
JHEP \textbf{09}, 176 (2020)
doi:10.1007/JHEP09(2020)176
[arXiv:2005.08990 [hep-th]].

\bibitem{Banerjee:2020vnt}
S.~Banerjee and S.~Ghosh,
``MHV Gluon Scattering Amplitudes from Celestial Current Algebras,'' JHEP \textbf{10} (2021), 111
[arXiv:2011.00017 [hep-th]].
\bibitem{Hu:2021lrx}
Y.~Hu, L.~Ren, A.~Y.~Srikant and A.~Volovich,
``Celestial dual superconformal symmetry, MHV amplitudes and differential equations,''
JHEP \textbf{12} (2021), 171
[arXiv:2106.16111 [hep-th]].
\bibitem{Pasterski:2017ylz}
S.~Pasterski, S.~H.~Shao and A.~Strominger,
``Gluon Amplitudes as 2d Conformal Correlators,''
Phys. Rev. D \textbf{96} (2017) no.8, 085006
doi:10.1103/PhysRevD.96.085006
[arXiv:1706.03917 [hep-th]].
\bibitem{Simmons-Duffin:2012juh}
D.~Simmons-Duffin,
``Projectors, Shadows, and Conformal Blocks,''
JHEP \textbf{04}, 146 (2014)
 doi:10.1007/JHEP04(2014)146
[arXiv:1204.3894 [hep-th]].
\bibitem{Dolan:2011dv}
F.~A.~Dolan and H.~Osborn,
``Conformal Partial Waves: Further Mathematical Results,''
[arXiv:1108.6194 [hep-th]].

\bibitem{Dolan:2000ut}
F.~A.~Dolan and H.~Osborn,
``Conformal four point functions and the operator product expansion,''
Nucl. Phys. B \textbf{599}, 459-496 (2001)
doi:10.1016/S0550-3213(01)00013-X
[arXiv:hep-th/0011040 [hep-th]].

\bibitem{Dolan:2003hv}
F.~A.~Dolan and H.~Osborn,
``Conformal partial waves and the operator product expansion,''
Nucl. Phys. B \textbf{678}, 491-507 (2004)
doi:10.1016/j.nuclphysb.2003.11.016
[arXiv:hep-th/0309180 [hep-th]].


\bibitem{Zamolodchikov:1984eqp}
A.~B.~Zamolodchikov,
``Conformal symmetry in two-dimensions: an explicit recurrence formula for the conformal partial wave amplitude,''
Commun. Math. Phys. \textbf{96}, 419-422 (1984)
doi:10.1007/BF01214585
[arXiv:1705.01027 [hep-th]].
\bibitem{Fan:2021isc}
W.~Fan, A.~Fotopoulos, S.~Stieberger, T.~R.~Taylor and B.~Zhu,
``Conformal blocks from celestial gluon amplitudes,''
JHEP \textbf{05}, 170 (2021)
doi:10.1007/JHEP05(2021)170
[arXiv:2103.04420 [hep-th]].
\bibitem{Fan:2021pbp}
W.~Fan, A.~Fotopoulos, S.~Stieberger, T.~R.~Taylor and B.~Zhu,
``Conformal blocks from celestial gluon amplitudes. Part II. Single-valued correlators,''
JHEP \textbf{11}, 179 (2021)
doi:10.1007/JHEP11(2021)179
[arXiv:2108.10337 [hep-th]].
\bibitem{Fan:2019emx}
W.~Fan, A.~Fotopoulos and T.~R.~Taylor,
``Soft Limits of Yang-Mills Amplitudes and Conformal Correlators,''
JHEP \textbf{05}, 121 (2019)
doi:10.1007/JHEP05(2019)121
[arXiv:1903.01676 [hep-th]].
\bibitem{Pate:2019lpp}
M.~Pate, A.~M.~Raclariu, A.~Strominger and E.~Y.~Yuan,
``Celestial operator products of gluons and gravitons,''
Rev. Math. Phys. \textbf{33}, no.09, 2140003 (2021)
doi:10.1142/S0129055X21400031
[arXiv:1910.07424 [hep-th]].
\bibitem{Alday:2009aq}
L.~F.~Alday, D.~Gaiotto and Y.~Tachikawa,
``Liouville Correlation Functions from Four-dimensional Gauge Theories,''
Lett. Math. Phys. \textbf{91}, 167-197 (2010)
doi:10.1007/s11005-010-0369-5
[arXiv:0906.3219 [hep-th]].
\bibitem{Nekrasov:2002qd}
N.~A.~Nekrasov,
``Seiberg-Witten prepotential from instanton counting,''
Adv. Theor. Math. Phys. \textbf{7}, no.5, 831-864 (2003)
doi:10.4310/ATMP.2003.v7.n5.a4
[arXiv:hep-th/0206161 [hep-th]].

\bibitem{Nekrasov:2003rj}
N.~Nekrasov and A.~Okounkov,
``Seiberg-Witten theory and random partitions,''
Prog. Math. \textbf{244}, 525-596 (2006)
doi:10.1007/0-8176-4467-9\_15
[arXiv:hep-th/0306238 [hep-th]].
\bibitem{Parke:1985pn}
S.~J.~Parke and T.~R.~Taylor,
``Perturbative QCD Utilizing Extended Supersymmetry,''
Phys. Lett. B \textbf{157}, 81 (1985)
[erratum: Phys. Lett. B \textbf{174}, 465 (1986)]
doi:10.1016/0370-2693(85)91216-X
\bibitem{Taylor:2017sph}
T.~R.~Taylor,
``A Course in Amplitudes,''
Phys. Rept. \textbf{691}, 1-37 (2017)
doi:10.1016/j.physrep.2017.05.002
[arXiv:1703.05670 [hep-th]].
\bibitem{deGioia:2022nkq}
L.~P.~de Gioia and A.~M.~Raclariu,
``Eikonal Approximation in Celestial CFT,''
[arXiv:2206.10547 [hep-th]].

\bibitem{Gonzo:2022tjm}
R.~Gonzo, T.~McLoughlin and A.~Puhm,
``Celestial holography on Kerr-Schild backgrounds,''
[arXiv:2207.13719 [hep-th]].



\end{thebibliography}
\end{document}